\documentstyle[aps,12pt]{revtex}

\begin{document}
\author{He-Shan Song\thanks{%
E-mail: hssong@dlut.edu.cn}, Dong Mi\thanks{%
E-mail: dongmi@dlut.edu.cn}, Ling Zhou and Chong Li}
\address{Department of Physics, Dalian University of Technology, \\
Dalian 116024, P. R. China}
\title{Quantum Entanglement of Photons in Doubled $q$-Fock Space}
\maketitle

\begin{abstract}
A doubled $q$-Fock space is constructed by introducing an idle mode system
dual to the physical one under consideration. The quantum entangtlements of
photons in the squeezed states and thermal states based on the douled $q$%
-Fock space are discussed.
\end{abstract}

\vskip 0.5cm

PACS numbers: 42.50.Dv, 42.50.Ar, 11.10.Wx

\baselineskip 21pt

{\bf 1. Introduction}

The concept of the quantum entanglement which can not be factorized into a
product of one-particle state was proposed by Schr\"{o}dinger in 1935 [1].
The quantum entanglement may be considered as most important and intriguing
character of quantum mechanics. In recent years, special attentions have
been paid to study quantum entangled states. Some quantum optical
experiments have already prepared quantum entangled states. A very
convenient method for preparing entangled photon pair by parametric
down-conversion in laser pumped non-linear crystal was discovered by Burnham
and Weinberg in 1970. Their discovery permitted the development of
two-photons interferometry by many authors[2]. By applying finite resolution
measurement, the non-classical correlations of entangled photons are
analyzed in the reference [3]. It has been demonstrated that the quantum
entanglement is key process in the quantum computation, quantum
teleportation, quantum cryptography and so on.

On the other hand, the study of exactly solvable statistical models has led
to a new algebra, which is deformation of universal enveloping algebra.
These deformed algebraic structures are now usually called quantum group
[4-6]. In recent years, considerable interest in mathematical physics has
concentrated on the quantum group and corresponding algebras [7-9]. Of
particular interest is the Bose realization of the quantum harmonic
oscillators, which play an important roles in the study of quantum group and
its applications in physics. For a system of $q$-deformed bosonic
oscillators (for a system of photons, for example), since the usual
commutation relations (Heisenberg-Weyl algebra) are replaced by the $q$%
-deformed one, the physical properties of system will be affected by the
deformation parameter $q$.

The purpose of present paper is to investigate some quantum properties such
as squeezing, quantum entanglement including thermal entanglement of photons
in the doubled $q$-Fock space. In the section 2, we will construct a doubled 
$q$-Fock space by introducing an idle mode system dual to the real one. In
section 3, we will discuss the entanglement of two-mode photons with
infinite degree of freedom in the doubled $q$-Fock space. In the section 4,
we will define a $q$-deformed thermal vacuum state and show the thermal
vacuum is in fact a thermally entangled state. Some conclusions are
summarized at the end of the paper.

{\bf 2. The doubled q-Fock space}

As is well known, an assembly of photons is simplest example of a Bose
system of independent particles. Photons are quanta of the electromagnetic
field. An electromagnetic field with frequency $\omega $ is equivalent to a
bosonic oscillators with frequency $\omega $ and the $n$-th excited level of
bosonic oscillator corresponds to the state of the electromagnetic field
with $n$-photons, each of which has the energy $\epsilon = 
\rlap{\protect\rule[1.1ex]{.325em}{.1ex}}h%
\omega $. So a state of an electromagnetic field can be denoted by the
number state $\left| n\right\rangle $ with $n=0,1,2,\cdots $. All states $%
\{\left| n\right\rangle $, $n=0,1,2,\cdots $. $\}$ construct a Fock space.
Different states in Fock space are connected by creation and annihilation
operators of photons.

Now we introduce operators $\{a,a^{+},N,1\}$ of $q$-deformed bosonic
oscillator and define, analogue to the usual bosonic oscillators, $q$%
-deformed Fock space [10]

\begin{equation}
\left\{ \left| n\right\rangle =\left( D_q(n)\right) ^{-\frac 12%
}(a^{+})^n\left| 0\right\rangle ,n=0,1,2,\cdots \ \right\}  \eqnum{2.1}
\end{equation}
based on the vacuum state defined as $a\left| 0\right\rangle =0,$ where $a$
denotes $q$-deformed annihilation operator of photon. The deformation
function $D_q(n)$ in the equation (2.1) satisfies

\begin{equation}
D_q(n)|_{q\rightarrow 1}=n,  \eqnum{2.2}
\end{equation}

\begin{equation}
D_q(0)=0,\text{ \quad }D_q(1)=1  \eqnum{2.3}
\end{equation}

and

\begin{equation}
D_q(n)!=\prod\limits_{k=1}^nD_q(k).  \eqnum{2.4}
\end{equation}

Analogue to the usual case, the action of annihilation (creation) operator $%
a $ ($a^{+}$) and number operator $N$ on the state $\left| n\right\rangle $
can be defined as

\begin{equation}
a\left| n\right\rangle =\sqrt{D_q(n)}\left| n-1\right\rangle ,  \eqnum{2.5}
\end{equation}

\begin{equation}
a^{+}\left| n\right\rangle =\sqrt{D_q(n+1)}\left| n+1\right\rangle , 
\eqnum{2.6}
\end{equation}

\begin{equation}
N\left| n\right\rangle =n\left| n\right\rangle .  \eqnum{2.7}
\end{equation}

Now we introduce an operator $P_{mn\text{ }}$defined as

\begin{equation}
P_{mn\text{ }}=\left| m\right\rangle \left\langle n\right|  \eqnum{2.8}
\end{equation}
with $m,n=0,1,2,\cdots $. The operator $P_{mn\text{ }}$ is an infinite
dimensional matrix in which the element at $m$-th row $n$-th column is one
and the others are zero. By using the operator $P_{mn\text{ }}$, one can
immediately find the matrix representations of the operators $%
\{a,a^{+},N,1\} $:

\begin{equation}
a=\sum\limits_{n=0}^\infty \sqrt{D_q(n+1)}P_{n,n+1},  \eqnum{2.9}
\end{equation}

\begin{equation}
a^{+}=\sum\limits_{n=0}^\infty \sqrt{D_q(n+1)}P_{n+1,n},  \eqnum{2.10}
\end{equation}

\begin{equation}
N=\sum\limits_{n=0}^\infty nP_{nn},  \eqnum{2.11}
\end{equation}

\begin{equation}
1=\sum\limits_{n=0}^\infty P_{nn}.  \eqnum{2.12}
\end{equation}
From equation (2.9) to equation (2.12) we find

\begin{equation}
a^{+}a=\sum\limits_{n=0}^\infty D_q(n)P_{nn}=D_q(N),  \eqnum{2.13}
\end{equation}

\begin{equation}
aa^{+}=\sum\limits_{n=0}^\infty D_q(n+1)P_{nn}=D_q(N+1)  \eqnum{2.14}
\end{equation}
and so we obtain the generalized $q$-Heisenberg-Weyl algebra

\begin{equation}
\left[ a,a^{+}\right] =D_q(N+1)-D_q(N),  \eqnum{2.15}
\end{equation}

\begin{equation}
\left[ N,a^{+}\right] =a^{+},  \eqnum{2.16}
\end{equation}

\begin{equation}
\left[ N,a\right] =-a.  \eqnum{2.17}
\end{equation}
Obviously,

\begin{equation}
D_q(N)\left| n\right\rangle =\sum\limits_{m=0}^\infty D_q(m)P_{mm}\left|
n\right\rangle =D_q(n)\left| n\right\rangle ,  \eqnum{2.18}
\end{equation}

\begin{equation}
D_q(N+1)\left| n\right\rangle =\sum\limits_{m=0}^\infty D_q(m+1)P_{mm}\left|
n\right\rangle =D_q(n+1)\left| n\right\rangle  \eqnum{2.19}
\end{equation}
and so the function $D_q(n)$ is eigenvalue of the operator $D_q(N)$.

Note that the deformation function $D_q(x)$ is not unique in general. The
form of $D_q(x)$ will be related with Bose realization scheme of $SU(2)$.
Different deformation functions will give different deformation algebras. It
is easy to show that deformation function $D_q(n)=\frac{q^n-q^{-n}}{q-q^{-1}}
$, for example, leads to the Biedenharn [6] and Mackfalan's [5] deformation
algebra

\begin{equation}
aa^{+}-qa^{+}a=q^{-N}.  \eqnum{2.20}
\end{equation}

Based on the $q$-Fock space, one can also construct doubled $q$-Fock space.
Since in quantum optics we often need consider two mode system such as
non-degenerate parametric amplifier gives rising to two mode squeezing and
the doubled Fock space has to be considered. In order to construct doubled $%
q $-Fock space, it is convenient to introduce an idle mode system which is
of exactly same structure as the physical one under consideration. We will
denote the quantities associated with the idle mode photons by tilde such as
idle photon number state by $\left| \widetilde{n}\right\rangle $,
annihilation (creation) operator of an idle photon by $\widetilde{a}$ $(%
\widetilde{a}^{+})$. Then the doubled $q$-Fock space spanned by direct
product $\left| n\right\rangle \otimes \left| \widetilde{n}\right\rangle
\equiv \left| n\widetilde{n}\right\rangle $ can be defined as

\begin{equation}
\left\{ \left| n\widetilde{n}\right\rangle =\frac{(a^{+})^n(\widetilde{a}%
^{+})^n}{D_q(n)!}\left| 0\widetilde{0}\right\rangle ,n=0,1,2,\cdots \right\}
,  \eqnum{2.21}
\end{equation}
where the tilde operators $\widetilde{a}$, $\widetilde{a}^{+}$ satisfy the
same algebra as corresponding non-tilde operators due to the duality of $q$%
-Fock space [11]. That is

\begin{equation}
\left[ \widetilde{a},\widetilde{a}^{+}\right] =D_q(\widetilde{N}+1)-D_q(%
\widetilde{N}),  \eqnum{2.22}
\end{equation}

\begin{equation}
\left[ \widetilde{N},\widetilde{a}^{+}\right] =\widetilde{a}^{+}, 
\eqnum{2.23}
\end{equation}

\begin{equation}
\left[ \widetilde{N},\widetilde{a}\right] =-\widetilde{a}.  \eqnum{2.24}
\end{equation}

In the doubled $q$-Fock space, the expectation value of a bose-like operator 
$F$ is given by

\begin{equation}
\left\langle \widetilde{m}n\right| F\left| n^{\prime }\widetilde{m}^{\prime
}\right\rangle =\left\langle n\right| F\left| n^{\prime }\right\rangle
\delta _{mm^{\prime }}  \eqnum{2.25}
\end{equation}
and that of corresponding $\widetilde{F}$ is

\begin{equation}
\left\langle \widetilde{m}n\right| \widetilde{F}\left| n^{\prime }\widetilde{%
m}^{\prime }\right\rangle =\left\langle \widetilde{m}\right| F\left| 
\widetilde{m}^{\prime }\right\rangle \delta _{nn^{\prime }}.  \eqnum{2.26}
\end{equation}
The actions of operators $\widetilde{a}$, $\widetilde{a}^{+}$ and $%
\widetilde{N}$ on the doubled $q$-Fock state $\left| n\widetilde{n}%
\right\rangle $ are defined, analogue to the equation (2.5) to equation
(2.7), as

\begin{equation}
\widetilde{a}\left| n\widetilde{n}\right\rangle =\sqrt{D_q(n)}\left| n,%
\widetilde{n}-1\right\rangle ,  \eqnum{2.27}
\end{equation}

\begin{equation}
\widetilde{a}^{+}\left| n\widetilde{n}\right\rangle =\sqrt{D_q(n+1)}\left| n,%
\widetilde{n}+1\right\rangle ,  \eqnum{2.28}
\end{equation}

\begin{equation}
\widetilde{N}\left| n\widetilde{n}\right\rangle =n\left| n\widetilde{n}%
\right\rangle .  \eqnum{2.29}
\end{equation}

Having defined the doubled $q$-Fock space, we can discuss some quantum
properties of photons in the doubled $q$-Fock space. In the subsequent
sections, we will investigate some quantum properties such as squeezing,
entanglement including thermal entanglement of photons in the doubled $q$%
-Fock space.

{\bf 3. Entanglement of photons in squeezed states}

Based on the above discussions, one can define squeezed vacuum state in the
doubled $q$-Fock space as

\begin{equation}
\left| O\left( \xi \right) \right\rangle =\exp \left\{ \xi \left( a^{+}%
\widetilde{a}^{+}-a\widetilde{a}\right) \right\} \left| 0\widetilde{0}%
\right\rangle =\sum\limits_{n=0}^\infty P_n^{\frac 12}\left| n\widetilde{n}%
\right\rangle ,  \eqnum{3.1}
\end{equation}
where $\left| n\widetilde{n}\right\rangle $ is given by the equation(2.21)
and

\begin{equation}
P_n=\frac{\tanh ^{2n}\xi }{\cosh ^2\xi }  \eqnum{3.2}
\end{equation}
denotes probability of $2n$-photons state. The squeezing parameter $\xi $ is
taken to be real for simplicity. It is easy to verify

\begin{equation}
\sum\limits_{n=0}^\infty P_n=\frac 1{\cosh ^2\xi }\sum\limits_{n=0}^\infty
\tanh ^{2n}\xi =1  \eqnum{3.3}
\end{equation}
and so the squeezed vacuum is normalized as

\begin{equation}
\left\langle O\left( \xi \right) \right| \left| O\left( \xi \right)
\right\rangle =\sum\limits_{m=0}^\infty \sum\limits_{n=0}^\infty P_m^{\frac 1%
2}P_n^{\frac 12}\left\langle n\widetilde{n}\right| \left| m\widetilde{m}%
\right\rangle =\sum\limits_{n=0}^\infty P_n=1.  \eqnum{3.4}
\end{equation}

From the equation (3.1) we see that the squeezed vacuum state $\left|
O\left( \xi \right) \right\rangle $ is an infinite superposition of doubled
number states $\left| n\widetilde{n}\right\rangle \equiv \left|
n\right\rangle \otimes \left| \widetilde{n}\right\rangle $ . The state $%
\left| O\left( \xi \right) \right\rangle $ can not be factorized into a
direct product of one-mode states. As photons are created in pair, there is
perfect correlation between the photons in each state. So we can regard the $%
q$-squeezed vacuum $\left| O\left( \xi \right) \right\rangle $ as an example
of partly entangled state for two-mode with $2n$-photons. Its entanglement
is given by Shannon's quantum entropy

\begin{eqnarray}
E &=&-\sum\limits_{n=0}^\infty P_n\log _2P_n  \nonumber \\
&=&-\sum\limits_{n=0}^\infty \frac{\tanh ^{2n}\xi }{\cosh ^2\xi }\ln \frac{%
\tanh ^{2n}\xi }{\cosh ^2\xi }  \nonumber \\
&=&-\left\{ \sinh ^2\xi \log _2\sinh ^2\xi -(1+\sinh ^2\xi )\log _2(1+\sinh
^2\xi )\right\}  \nonumber \\
&=&-\left\{ \overline{n_0}\log _2\overline{n_0}-(1+\overline{n_0})\log _2(1+%
\overline{n_0})\right\} ,  \eqnum{3.5}
\end{eqnarray}
where $\overline{n_0}=\sinh ^2\xi $ denotes the average number of photons in
the undeformed squeezed state, and $k$ is Boltzmann constant.

The average number of photons in the $q$-squeezed vacuum is given by

\begin{equation}
\overline{n}=\left\langle O\left( \xi \right) \right| a^{+}a\left| O\left(
\xi \right) \right\rangle =\sum\limits_{n=0}^\infty D_q(n)P_n.  \eqnum{3.6}
\end{equation}
If we take the Biedenharn and Mackfalan's deformation scheme

\begin{equation}
D_q(n)=\frac{q^n-q^{-n}}{q-q^{-1}}  \eqnum{3.7}
\end{equation}
and substituting $D_q(n)$ into the equation (3.6), we obtain

\begin{equation}
\overline{n}=\frac{\tanh ^2\xi }{\cosh ^2\xi }\left\{ \frac{C_1}{1-q\tanh
^2\xi }+\frac{C_2}{1-q^{-1}\tanh ^2\xi }\right\}  \eqnum{3.8}
\end{equation}
with

\begin{equation}
C_1=\frac q{q-q^{-1}},\text{ \quad }C_2=\frac{q^{-1}}{q^{-1}-q},  \eqnum{3.9}
\end{equation}

\begin{equation}
C_1\ +C_2=1\ .  \eqnum{3.10}
\end{equation}
It is easy to see that when $q\rightarrow 1$, the equation (3.8) reduces to $%
\overline{n}=\sinh ^2\xi ,$ which is just the average number of photons in
undeformed squeezed state.

We see from above discussions that the entanglement of two-photon is
dependent only on the parameter $\xi $ but independent of the deformation
parameter $q$. The average photon number, however, depended on both of
parameters $\xi $ and $q$ as seen from equation (3.8). The average number of
photons deviated from $\sinh ^2\xi $ and divided into two parts due to $q$%
-deformation of the Fock space.

Consider now the quantum fluctuations of two quadratures $U_1$ and $U_2$.
The quadratures are defined as 
\begin{equation}
U_1\ =(a+a^{+}+\widetilde{a}+\widetilde{a}^{+})/2^{\frac 32}\ ,  \eqnum{3.11}
\end{equation}

\begin{equation}
U_2\ =(a-a^{+}+\widetilde{a}-\widetilde{a}^{+})/2^{\frac 32}i.  \eqnum{3.12}
\end{equation}
Then the quantum fluctuations of $U_i$ $(i=1,2)$ in the $q$-squeezed vacuum
are given by

\begin{equation}
\left( \Delta U_i\right) ^2=\left\langle O\left( \xi \right) \right|
U_i^2\left| O\left( \xi \right) \right\rangle -\left( \left\langle O\left(
\xi \right) \right| U_i\left| O\left( \xi \right) \right\rangle \right) ^2. 
\eqnum{3.13}
\end{equation}
It is easy to calculate the vacuum expectation values

\begin{equation}
\left\langle O\left( \xi \right) \right| a^{+}a\left| O\left( \xi \right)
\right\rangle =\overline{n},  \eqnum{3.14 }  \label{ }
\end{equation}

\begin{equation}
\left\langle O\left( \xi \right) \right| aa^{+}\left| O\left( \xi \right)
\right\rangle =\overline{n}/\tanh ^2\xi ,  \eqnum{3.15}
\end{equation}

\begin{equation}
\left\langle O\left( \xi \right) \right| a\widetilde{a}\left| O\left( \xi
\right) \right\rangle =\overline{n}/\tanh \xi ,  \eqnum{3.16}
\end{equation}

\begin{equation}
\left\langle O\left( \xi \right) \right| \widetilde{a}\widetilde{a}%
^{+}\left| O\left( \xi \right) \right\rangle =\overline{n}/\tanh \xi 
\eqnum{3.17}
\end{equation}
and the others are zero. Substituting these results to the equation (3.13)
we get

\begin{equation}
\left( \Delta U_1\right) ^2=\frac 14\frac{\left( 1+\tanh \xi \right) ^2}{%
\tanh ^2\xi \ }\overline{n}\ ,  \eqnum{3.18}
\end{equation}

\begin{equation}
\left( \Delta U_2\right) ^2=\frac 14\frac{\left( 1-\tanh \xi \right) ^2}{%
\tanh ^2\xi \ }\overline{n}\   \eqnum{3.19}
\end{equation}
and the corresponding uncertainty relation gives

\begin{equation}
\left( \Delta U_1\right) ^2\cdot \left( \Delta U_2\right) ^2=\ \left( \frac{%
\overline{n}}{4\sinh ^2\xi \ }\right) ^2  \eqnum{3.20}
\end{equation}
with $\overline{n}$ equal to the equation (3.8).

From equation (3.18) to (3.20) we see that when $q\rightarrow 1$,

\begin{equation}
\left( \Delta U_1\right) ^2=\frac 14\frac{\left( 1+\tanh \xi \right) ^2}{%
\tanh ^2\xi \ }\sinh ^2\xi =\frac 14e^{2\xi }\ ,  \eqnum{3.21}
\end{equation}

\begin{equation}
\left( \Delta U_2\right) ^2=\frac 14\frac{\left( 1-\tanh \xi \right) ^2}{%
\tanh ^2\xi \ }\sinh ^2\xi \ =\frac 14e^{-2\xi },  \eqnum{3.22}
\end{equation}
and

\begin{equation}
\left( \Delta U_1\right) ^2\cdot \left( \Delta U_2\right) ^2=\frac 1{16}. 
\eqnum{3.23}
\end{equation}
So the fluctuations of two quadrature reduced to minimum uncertainty
relation as expected.

{\bf 4. Thermal entanglement of photons}

One can also discuss the properties of photons in the $q$-deformed thermal
states. To achieve this, a $q$-deformation of the thermal state have to be
developed. The crucial point is to define $q$-deformed vacuum $\left|
O\left( \xi \right) \right\rangle $. The $q$-deformed thermal vacuum $\left|
O\left( \xi \right) \right\rangle $ can be defined as [9]

\begin{equation}
\left| O\left( \beta \right) \right\rangle =\sum\limits_{n=0}^\infty P_n^{%
\frac 12}\left| n\widetilde{n}\right\rangle ,  \eqnum{4.1}
\end{equation}
where the doubled $q$-Fock state $\left| n\widetilde{n}\right\rangle $ is
given by the equation (2.22) and

\begin{equation}
P_n=\frac{e^{-\beta E_n}}{Z(\beta )}  \eqnum{4.2}
\end{equation}
denotes probability of the state with energy $E_n$ and $Z(\beta )$ is the
partition function. The thermal vacuum is normalized as

\begin{equation}
\left\langle O\left( \beta \right) \right| \left| O\left( \beta \right)
\right\rangle =\sum\limits_{m=0}^\infty \sum\limits_{n=0}^\infty P_m^{\frac 1%
2}P_n^{\frac 12}\left\langle m\widetilde{m}\right| \left| n\widetilde{n}%
\right\rangle =\sum\limits_{n=0}^\infty P_n=1.  \eqnum{4.3}
\end{equation}

Once that the $q$-thermal vacuum is defined, we can define further $q$%
-thermal creation and annihilation operators through Bogoliubov
transformation. By using the creation and annihilation operators on the $q$%
-thermal vacuum, a complete orthonormal set of state vectors containing $%
\left| O\left( \xi \right) \right\rangle $ as one of the members can finally
be found. This process is completely parallel to the corresponding process
in undeformed thermal field dynamics [11].

Consider now the average photon number in the $q$-thermal vacuum. Analogue
to the usual case, the average photon number can be found as

\begin{equation}
\overline{n}=\left\langle O\left( \beta \right) \right| a^{+}a\left| O\left(
\beta \right) \right\rangle .  \eqnum{4.4}
\end{equation}
Note that the operators $a,a^{+}$ in the equation (4.4) satisfy the $q$%
-deformed commutation relation, the equation (2.20) in the Biedemharn and
Mackfalan's deformation scheme.

For the photons (bosons) system, the thermal vacuum $\left| O\left( \beta
\right) \right\rangle $ is given by equation (4.1) with probability

\begin{equation}
\ P_n=\frac{\exp (-\beta E_n)}{Tr(e^{-\beta H})}=[1-\exp (-\beta \omega
)]\exp (-n\beta \omega ).  \eqnum{4.5}
\end{equation}
In the equation (4.5), $\beta =\frac 1{kT}$ with Boltzmann constant $k$ and
temperature $T$. The free Hamiltonian $H$ of the system is given by [7]

\begin{equation}
H=\omega N.  \eqnum{4.6}
\end{equation}
Substituting equation (4.5) into the equation (4.4) and by using equation
(2.18) and $D_q(n)$ in the Biedonharn and Mackfalan's scheme we get

$\ $%
\begin{equation}
\overline{n}=\ \left\{ \frac{C_1}{\exp \{\beta (\omega +\frac \lambda \beta
)\}-1}+\frac{C_2}{\exp \{\beta (\omega -\frac \lambda \beta )\}-1\ }\right\}
\eqnum{4.7}
\end{equation}
with $\lambda =\ln q$ and

\begin{equation}
C_1=\frac{q-1}{q-q^{-1}},\text{ \quad }C_2=\frac{1-q^{-1}}{q\ -q^{-1}}. 
\eqnum{4.8}
\end{equation}

\begin{equation}
C_1+C_2=1.  \eqnum{4.9}
\end{equation}
We see from the equation (4.7) that the distribution function (average
photon number) of the photons now divided into two part: one with energy $%
\omega +\frac \lambda \beta $ and the other with energy $\omega -\frac %
\lambda \beta $ due to the $q$-deformation of commutation relation. In the
equation (4.7), in order to guarantees the weights $C_1,C_2>0$ and $%
C_1+C_2=1 $, $q>0$ has to be assumed.

It is easy to check that when $q\rightarrow 1$, the photon number
distribution turn out to be

\begin{equation}
\overline{n}\ =\frac{\ 1}{\exp (\beta \ \omega )-1\ }\   \eqnum{4.10}
\end{equation}
as we expected.

Analogue to the section 3, one can also discuss the fluctuations of two
quadrature in thermal state. The fluctuations are given by

\begin{equation}
\left( \Delta U_i\right) ^2=\left\langle O\left( \beta \right) \right|
U_i^2\left| O\left( \beta \right) \right\rangle -\left( \left\langle O\left(
\beta \right) \right| U_i\left| O\left( \beta \right) \right\rangle \right)
^2.  \eqnum{4.11}
\end{equation}
Straightforward calculations gives the expectation values

\begin{equation}
\left\langle O\left( \beta \right) \right| a^{+}a\left| O\left( \beta
\right) \right\rangle =\overline{n},  \eqnum{4.12}
\end{equation}

\begin{equation}
\left\langle O\left( \beta \right) \right| aa^{+}\left| O\left( \beta
\right) \right\rangle =e^{-\beta \omega }\cdot \overline{n}\ ,  \eqnum{4.13}
\end{equation}

\begin{equation}
\left\langle O\left( \beta \right) \right| a\widetilde{a}\left| O\left(
\beta \right) \right\rangle =e^{-\frac 12\beta \omega }\cdot \overline{n}\ ,
\eqnum{4.14}
\end{equation}

\begin{equation}
\left\langle O\left( \beta \right) \right| a^{+}\widetilde{a}^{+}\left|
O\left( \beta \right) \right\rangle =e^{-\frac 12\beta \omega }\cdot 
\overline{n}\   \eqnum{4.15}
\end{equation}
and the others are zero. Substituting these results to the equation (4.11)
we obtain

\begin{equation}
\left( \Delta U_1\right) ^2=\frac 14\left( e^{\frac 12\beta \omega }+1\
\right) ^2\overline{n}\ ,  \eqnum{4.16}
\end{equation}

\begin{equation}
\left( \Delta U_2\right) ^2=\frac 14\left( e^{\frac 12\beta \omega }-1\
\right) ^2\overline{n}\ ,\   \eqnum{4.17}
\end{equation}
and the corresponding uncertainty relation gives

\begin{equation}
\left( \Delta U_1\right) ^2\cdot \left( \Delta U_2\right) ^2=\frac 1{16}\
\left( e^{\beta \omega }-1\ \right) ^2\overline{n}^2.  \eqnum{4.18}
\end{equation}
when $q\rightarrow 1$, the average number of photon $\overline{n}\rightarrow 
$ $\frac{\ 1}{\exp (\beta \ \omega )-1\ }$ and

\begin{equation}
\left( \Delta U_1\right) ^2|_{q\rightarrow 1}=\frac 14\frac{e^{\frac 12\beta
\omega }+1}{e^{\frac 12\beta \omega }-1},  \eqnum{4.19}
\end{equation}

\begin{equation}
\left( \Delta U_2\right) ^2|_{q\rightarrow 1}=\frac 14\frac{e^{\frac 12\beta
\omega }-1}{e^{\frac 12\beta \omega }+1}.  \eqnum{4.20}
\end{equation}
Thus we find

\begin{equation}
\left( \Delta U_1\right) ^2\cdot \left( \Delta U_2\right) ^2|_{q\rightarrow
1}=\frac 1{16},  \eqnum{4.21}
\end{equation}
namely, the fluctuations satisfy minimum uncertainty relation.

From above discussions we see that the behavior of photons in the $q$%
-thermal state is completely same with the behavior of photons in the $q$%
-squeezed states. The only difference is the factor $tanh\xi $ in the $q$%
-squeezed system is displaced by the factor $e^{-\frac 12\beta \omega }$ in
the thermal system. The thermal vacuum defined in equation (4.1) is an
infinite superposition of doubled number state $\left| n\widetilde{n}%
\right\rangle $ constructed by direct product of single-mode states $\left|
n\right\rangle $ and $\left| \widetilde{n}\right\rangle $. So we conclude
that the thermal vacuum is also a kind of entangled state, the thermally
entangled state. The entanglement of the system can be found as

\begin{eqnarray}
\ E &=&-\sum\limits_{n=0}^\infty P_n\log _2P_n  \nonumber \\
&=&-[1-\exp (-\beta \omega )]\exp (-n\beta \omega )\log _2\left\{ [1-\exp
(-\beta \omega )]\exp (-n\beta \omega )\right\}  \nonumber \\
&=&-\left\{ \overline{n_0}\log _2\overline{n_0}-(1+\overline{n_0})\log _2(1+%
\overline{n_0})\right\} ,  \eqnum{4.22}
\end{eqnarray}
where $\overline{n_0}=$ $\frac{\ 1}{\exp (\beta \ \omega )-1\ }$ denotes the
average number of photons in the undeformed thermal state. We see that the
entanglement of photons is depended on the parameter $\beta =\frac 1{kT}$,
the temperature of the system.

{\large ~}{\bf 5. Conclusions and Discussions }

We constructed a doubled $q$-Fock space by introducing an idle system dual
to the physical one under consideration and investigated the effects of $q$%
-deformation of Heisenberg-Weyl algebras to the average numbers and quantum
fluctuations in the squeezed state and thermal state. The results show that
the $q$-deformation decompound the photon distribution into two parts and
leads to energy shift $\pm \frac \gamma \beta $ in the thermal state. We
pointed out that the new vacuums, the squeezed vacuum $\left| O\left( \xi
\right) \right\rangle $ and thermal vacuum $\left| O\left( \beta \right)
\right\rangle $ in the doubled q-Fock space are entangled states of two-mode
with infinite degree of freedoms.

It is worthy to discuss that what is the meaning of the idle (tilde) system
and why we must construct a doubled space in order to define the new vacuums
such as squeezed vacuum $\left| O\left( \xi \right) \right\rangle $ or
thermal vacuum $\left| O\left( \beta \right) \right\rangle $. We will
discuss these questions in the forthcoming works [12].

\noindent {\LARGE Acknowledgments}

This project is supported by the Chinese Education  Foundation through grant
no. 1999014105.

\end{document}